\documentclass[a4paper,9pt]{article}

\usepackage{anysize}
\marginsize{50pt}{40pt}{0pt}{60pt}
\usepackage{placeins}
\usepackage[version=4]{mhchem}
\usepackage{siunitx}
\DeclareSIUnit\Molar{M}
\usepackage[utf8]{inputenc}
\usepackage{graphicx}
\usepackage[dvipsnames]{xcolor}
\usepackage{array}
\usepackage{multirow}
\usepackage{float}
\usepackage{amsmath}
\usepackage{amsfonts}
\usepackage{amssymb} 
\usepackage{bbm}
\usepackage{microtype}
\usepackage{booktabs}
\usepackage{verbatim}

\definecolor{CC_blue}{RGB}{16, 38, 148}  
\usepackage[colorlinks=true,
            urlcolor=CC_blue,
            citecolor=CC_blue,
            linkcolor=CC_blue,
            bookmarks=true,
            breaklinks]{hyperref}

\def\sectionautorefname{Section}

\usepackage[                
        backend=bibtex,      
        style=numeric,      
        natbib=true,        
        hyperref=true,      
        alldates=year,      
	    uniquelist=false,   
        sorting=none,
	    isbn=false,
	    eprint=false,
	    doi=false,
        url=false,
	    maxcitenames=2,
        giveninits=true
]{biblatex}

\renewbibmacro{in:}{%
  \ifentrytype{article}{}{%
    \printtext{\bibstring{in}\intitlepunct}%
  }
}

\AtEveryBibitem{
    \clearfield{urlyear}
    \clearfield{urlmonth}
    \clearlist{language}
}

\begin{document}
\twocolumn[{\centering{\Huge Continuous benchmarking: Keeping pace with an evolving ecosystem of models and technologies \par}\vspace{3ex}
	{\Large 
    Jan Vogelsang\,$^{1,7,\dag}$,
    Melissa Lober\,$^{2,7}$,
    Catherine Mia Schöfmann\,$^{1,7}$,
    José Villamar\,$^{2,7}$,
    Dennis Terhorst\,$^{2}$,
    Johanna Senk\,$^{2,3}$,
    Hans Ekkehard Plesser\,$^{2,4}$,
    Markus Diesmann\,$^{2,5,6}$,
    Susanne Kunkel\,$^{1}$,
    Anno C. Kurth\,$^{2, 8}$\par}\vspace{2ex}
    \small
    $^{1}$ Neuromorphic Software Ecosystems (PGI-15), Jülich Research Centre, Jülich, Germany\\
    $^{2}$ Institute for Advanced Simulation (IAS-6), Jülich Research Centre, Jülich, Germany\\
    $^{3}$ Sussex AI, School of Engineering and Informatics, University of Sussex, Brighton, United Kingdom\\
    $^{4}$ Faculty of Science and Technology, Norwegian University of Life Sciences, Ås, Norway\\
    $^{5}$ JARA-Institute Brain Structure-Function Relationships (INM-10), Jülich Research Centre, Jülich, Germany\\
    $^{6}$ Department of Physics, Faculty 1 \& Department of Psychiatry, Psychotherapy, and Psychosomatics, Medical School, RWTH Aachen University, Aachen, Germany\\
    $^{7}$ RWTH Aachen University, Aachen, Germany\\
    $^{8}$ Current address: Hierarchical Neural Computation RIKEN ECL Research Unit, RIKEN Center for Brain Science, Wako, Japan\\ 
    \vspace{2ex} 
    \textsuperscript{\dag} \href{mailto:j.vogelsang@fz-juelich.de}{j.vogelsang@fz-juelich.de}\\
    \vspace{2ex}}

\noindent
Drawing on ideas from continuous integration, we present concepts of an automated benchmarking pipeline for high performance applications. Customization and collaboration have been key design goals owing to the requirements of research-software development as a continuous community effort. We have extended our previous conceptual work on systematic benchmarking workflows with the functionality of user-agnostic operations as well as continuous benchmarking. This fosters reproducibility and re-use of benchmarking results to ensure sustainable technological progress. We provide software-engineering solutions to keep pace with the rapid evolution of both large-scale models and high-performance computing systems with a view towards the scientific domains of neuroscience and artificial intelligence.

\medbreak
\par\vspace{5ex}]

\section{Introduction}
\label{sec:introduction}
%
Over the last decades, software has become ever more essential for research. Despite its ubiquity, its importance often remains overlooked \citep{Hocquet24_1}, even though results from computational science inform public policy on a large societal scale, as exemplified by epidemiological modeling of COVID infection dynamics~\citep{Ferguson20, Adam20_316} and climate simulations~\citep{Lee23}.

The emerging field of \textit{Research Software Engineering}~\citep{Baxter12, Speck24_8} aims at addressing the specific challenges encountered in writing, extending, and maintaining scientific software. Software project management distinguishes the verification and validation of scientific software \citep{Oberkampf10_verification}. Verification is concerned with the correctness of the software. In large collaborative software projects the standard approach ensuring this is automated testing, for example in the form of \textit{continuous integration}~(CI) \citep{Shahin17_3903}.
Validation, on the other hand, is about meeting the needs of the users.
Certain scientific questions critically depend on the size of the studied systems~\citep{Anderson72_393, Albada15}, and thus can only be investigated using high-performance computing (HPC). At the same time, also the size of experimentally collected data can, in certain fields, cross the peta-byte boundary requiring highly data intensive and computationally demanding analyses. Speed of code execution, memory requirements as well as data processing limitations are therefore part of the validation of scientific software as they limit the range of possible scientific questions that can be addressed.

In the case of HPC applications such as simulation engines (dedicated scientific software computing numerical solutions for the mathematical formulation of certain natural phenomena), validation thus requires rigorous performance benchmarking, that is the evaluation of metrics (for example, time-to-solution, memory footprint, energy consumption, etc.) which allow researchers to assess the performance of a simulation engine across various configurations such as hardware resources, build specifications, or model parameter spaces. This aspect is often not addressed systematically and comes with its own, unique difficulties. \citet{Albers22_837549} analyze the situation and identify five dimensions characterizing the complexity of HPC benchmarking: ``Hardware configuration'', ``Software configuration'', ``Simulators'', ``Models and parameters'', and ``Researcher communication''. The authors argue for the need of a common benchmarking framework to support the reliability, reproducibility, and comparability of performance benchmarks. Here, ``reproducibility'' refers to the ability of an independent group of researchers to obtain the same results as a reference group when using the same experimental setup\footnote{\url{https://www.acm.org/publications/policies/artifact-review-and-badging-current}}. It is distinguished from ``repeatability'' (same group of researchers obtaining the same results with the same setup) and ``replicability'' (independent group using different setup). 
\citet{Albers22_837549} propose to employ an array of standard benchmark models together with a standardized way for running them. This crucially includes all steps from configuration via execution to handling of results, as well as tracking of accruing metadata. The outcome of their work is a conceptual benchmarking workflow, distinguishing, for example, model, user, and machine configuration. The accompanying reference implementation \texttt{beNNch}\footnote{\url{https://github.com/INM-6/beNNch}} is geared towards benchmarking spiking neural network simulation engines and provides a modular implementation of the conceptual workflow, aiming to establish benchmarking as a collaborative, routine effort. Since its inception \texttt{beNNch} has been aiding the development of the large-scale neural network simulation code \texttt{NEST} \citep{Gewaltig_07_11204}.

In practice, beyond its initial use, \texttt{beNNch} did not accomplish its goals.
\citet{Villamar25_942} already identify significant shortcomings and propose more comprehensive practices for collecting the full ensemble of information required to document the entire workflow execution.
Although \texttt{beNNch} standardizes the benchmark execution phase, the underlying environment configuration remains in the responsibility of the individual researcher, and the sensitive interplay between various hardware and software configurations in HPC systems amplifies minute differences. Taken together, this makes replicating, or even reproducing benchmarking results a challenging task. As a consequence, extensive expert knowledge regarding software and hardware configurations is required from all researchers using \texttt{beNNch}.

%
%
%

The COVID-19 pandemic provided for an unexpected stress test of \texttt{beNNch} as a tool for reliable benchmarking based on recorded specifications. Much of the work on the \texttt{beNNch} framework was ready for use before COVID-19 struck in the spring of 2020, even though the publication about it did not appear in preprint form before December 2021 and as a formal paper in May 2022 \citep{Albers22_837549}. In March 2020, the pandemic forced staff at all four laboratories involved in the development of \texttt{beNNch} into remote work for much of the next two years. In this situation, researchers from J\"ulich Research Centre involved in \texttt{beNNch} attempted to use \texttt{beNNch} to reproduce existing and implement new benchmarks. In several cases, participating researchers obtained significantly different results using supposedly identical \texttt{beNNch} specifications. With communication limited to emails, group chats and video conferences, it took in some cases months to identify minute discrepancies in configurations that explained these differences. While we have no hard proof, based on previous experience we consider it highly likely that these issues would have been resolved much faster in a pre-pandemic in-person work environment with a low threshold for pulling all relevant colleagues into brainstorming meetings on short notice \cite{Vier_2022_331}. Unfortunately, it appears that even several years after all lock-downs were lifted, many work environments have still not returned to the same level of spontaneous on-site interaction that many of us remember from before 2020. An important conclusion from this experience is that \citet{Albers22_837549} underestimated the communication challenges facing complex benchmark experiments: While they believed that researchers would face communication challenges mainly in the final data-presentation stage of the \texttt{beNNch} workflow, the pandemic experience revealed key challenges in the communication of complex configuration information between researchers. This indicates a need for robust digital support for configuration management in benchmarking workflows.

\begin{figure*}
\begin{center}
\includegraphics[width=0.75\textwidth]{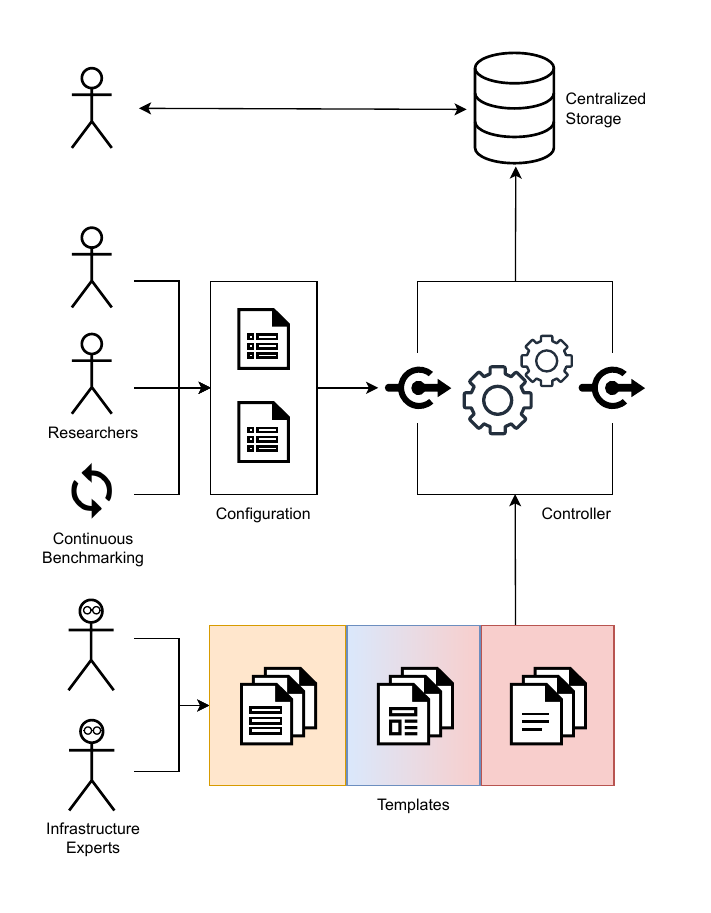}
\end{center}
 \caption{
Overview of the continuous benchmarking process. Researchers specify their experiments via configurations and use them to trigger benchmarking runs through a unified entry point (middle row). A researcher can also query data and metadata from previous runs through a centralized storage (top row). Infrastructure experts (for example system administrators or senior researchers; bottom row) define the hierarchical templates consisting of workflow (orange), infrastructure (blue: platform, red: machine), and implementation (red) templates, which are then used by a controller to construct a benchmarking pipeline given the researcher’s configuration and execute it to produce benchmark results.
 }
\label{fig:bench_concept}
\end{figure*}

Building on core insights from continuous integration \citep{nuyujukian23_dev_ops}, we here propose an improved point of view as well as practices on the previously developed performance benchmarking workflow: An  entry point agnostic of researcher and machine removes the need for individual hardware and software configuration by adding an additional level of abstraction that facilitates automated building and testing (see \autoref{fig:bench_concept} for a graphical overview). This not only ensures clear software setups but also identical execution of the different steps on HPC systems. Additionally, this approach enables \textit{continuous benchmarking}~(CB): the regular assessment of the performance of an HPC application, akin to the automated testing practice in continuous integration. This assessment must represent a wide range of use cases of the HPC application, and thereby allow researchers to identify formally correct changes to the code that nonetheless negatively impact the simulation engine's performance. Additionally, the added level of abstraction facilitates a separation of responsibilities. Through the specification of a common benchmarking setup by specialists, we reduce the required domain knowledge of each individual contributor.

The recent EuroHPC joint undertaking~\citep{eurohpc} states increasing CI support at HPC facilities as one of its major goals, underlining the importance of CI for HPC software development.

The feasibility of integrating CB into the CI framework of a scientific software has previously been shown by \citet{Anzt19_1}. Their solution is based on a predefined pipeline which is setup only once. This design prevents running benchmarks dynamically \textemdash changing benchmark parameters on demand that differ from the predefined battery of benchmarks. On the other hand, \citet{Pearce23} developed a CI-based benchmarking framework that targets HPC-systems benchmarks. Recently, \citet{exaCB} developed a CI-based benchmarking solution targeting HPC systems and parallel applications. Similarly, to \citet{Anzt19_1} their design is based on a pre-defined pipeline which introduces challenges when deploying their architecture across machines.

The solution proposed here, which we name \texttt{CI-beNNch} (\autoref{fig:bench_concept}), focuses on HPC-based applications. In contrast to the work of \citet{Anzt19_1}, \citet{Pearce23}, and \citet{exaCB}, we strive for greater generality and a higher level of abstraction by addressing underlying issues that complicate performance benchmarking on HPC systems, increasing accessibility. The increased generality additionally widens the scope of potential HPC applications beyond benchmarking. However, for validation of our solution, we focus on the continuous benchmarking use case.

In the next section, we discuss difficulties encountered when using \texttt{beNNch} and identify requirements for a comprehensive benchmarking setup, not offered by existing solutions. Subsequently, we present generic use cases for the improved scheme, highlighting its flexibility. Finally, we discuss how the scheme derived here can be applied beyond CB, choosing as a specific use case the development of new features for the \texttt{NEST} simulation code.

Preliminary results have been presented in abstract form \citep{nest-conf-24}.

\section{Removing the individual from the equation}
\label{sec:methods}
The goal of any benchmarking framework must be to achieve repeatability, replicability, and reproducibility independent of the individual who composed the benchmark and the infrastructure used for benchmark execution. In the following subsection, we highlight hindrances for achieving this goal. Our assessment is structured along shortcomings and problems encountered during three years of practical application of \texttt{beNNch}. Subsequently, we offer systemic solutions to these problems.

\subsection{Hindrances for repeatability, replicability, and reproducibility}
In \texttt{beNNch}, repeatability is ensured by identically executing workflows given the same benchmark configuration. Clearly, this is only possible if the experimental setup stays identical, including the executing hardware. As HPC hard- and software undergoes frequent changes, an identical experimental setup cannot be guaranteed in general. While it is impossible to avoid changing setups, in order to reduce the risk of errors, a benchmarking framework should minimize the work required to achieve the most similar benchmarking setup possible. Since \texttt{beNNch} does not manage software environments, this task is in the responsibility of the individual researcher. A changing machine therefore requires researchers to replicate the environment. This often leads to entirely different work steps compared to the original setup, in turn leading to even more disparate environments than necessary.
Further, individual and private environments hinder reproducibility: another researcher trying to reproduce a benchmark must setup an individual environment as well, which results in no guarantees of running an identical benchmark. Additionally, while a central storage location for base configurations exists, production use has shown that these base configurations were typically overridden by researchers on the target machine, and continuously changed over the course of multiple benchmarks. Other researchers trying to reproduce a benchmark are therefore not able to directly access the configuration without communicating with the relevant researcher, which can be time-consuming and becomes particularly challenging when trying to replicate data from a while back.
These problems are further amplified when trying to replicate results using a different experimental setup. As \texttt{beNNch} has to be deployed on each machine individually, it did not provide the functionality to port a benchmark to another system and run it under comparable conditions. Even though HPC machines typically only differ in some details (such as the optimal scheduler parameters, dependency module names or versions, and storage locations), moving to a new machine not only requires to recreate a suitable  environment, but also to redo any machine specific changes to \texttt{beNNch}. While these changes could in theory be shared among researchers, changing the tool itself to support additional machines is not good practice.
Consequently, the current workflow does in practice not fully support repeatability, replicability, and reproducibility. Expert knowledge required at each step is the main issue during routine use of \texttt{beNNch}. Extensive prior knowledge is required to execute a benchmark, as each researcher must manage individual environments, setup and adapt the tool based on the machine, use it correctly, and configure the benchmark's hardware, software, model, and simulator parameters. This turned out to be a major challenge, not only for junior researchers which often have limited HPC knowledge, but also for senior staff which had to first learn the intricacies of \texttt{beNNch} to run benchmarks.

\subsection{Overcoming the hindrances}
To solve these apparent issues, we propose the integration of benchmarking suites into CI setups with a unified entry point from which all benchmarks are triggered. Thereby, the execution of the benchmark becomes independent of the researcher, unwanted side effects caused by different environments and benchmark execution paths can be largely avoided, a central location where benchmark results of involved researchers of a research group are stored can be specified, and an output channel for automatically generated default analyses can be set up.

\paragraph{Standardized and centralized hierarchical configurations}
In order to facilitate such an integration, we suggest hierarchical benchmark configurations. Adapting another researcher's specification based on the personal restrictions given by the soft- and hardware environment can be achieved by overwriting individual parts of the given benchmark specification and inheriting all unmodified values. For a specific research group, a typical approach becomes to identify the set of all parameters researchers should be able to specify and create a base configuration for each common benchmark used in the group. Additionally, and crucially, researchers are not required to be equally knowledgeable in all domains, but can focus on their specific questions while relying on experts for remaining setups to be properly configured. Finally, configurations are easily exchangeable between different research groups.

The configuration defines everything required to execute a benchmark. For example, it not only specifies software dependencies, but also provides the actual way of fulfilling them, e.g. by loading the corresponding module on an HPC system. Ideally, the benchmarking workflow should not require any user interaction other than providing the benchmark configuration as input and receiving measurement data from the benchmark execution as output. In \texttt{beNNch}, there was no guarantee that a configuration authored by one researcher could be successfully used by another researcher straight away. Treating the benchmarking pipeline as a self-contained service effectively removes the dependence on the person executing the benchmark, maximizing reproducibility and further reducing the initial expert knowledge needed to execute benchmarks. All benchmark configurations must therefore be stored in a shared location. In this way, they can be accessed by as many other researchers as possible, and are thus not hidden behind individual user accounts. Thus, any researcher is now able to run and modify a specific benchmark. This follows the idea of standard CI workflows, where tests and builds are typically executed by a cloud runner without requiring any human intervention \citep{Zhao17_60}.

A benchmark defined by a standardized configuration ensures independence of the benchmarking framework implementation and makes it agnostic of the platform on which the benchmark is executed. In this context, we define the platform as the CI framework, responsible for storing and running the benchmarking framework. A common situation for HPC systems are GitLab CI pipelines\footnote{\url{https://gitlab.com/rluna-gitlab/gitlab-ce}} running, for example, on a dedicated system provided by the maintainer of the supercomputer. However, while the CI tool itself may be standard, the implementation of the benchmarking workflow varies significantly across different HPC infrastructures. Challenges such as heterogeneous architectures, job scheduling latencies, and security barriers require unique workflow adaptations for each system: if on an HPC system there are different types of compute nodes, it is important to compile the simulator and execute the simulation on the same node type; on systems with resource managers scheduling compute jobs once resources become available, the benchmarking workflow must create jobs accordingly and wait for scheduled jobs to be finished; on systems restricting internet access for compute nodes required software must be downloaded in a separate step on a dedicated node and subsequently transferred to the compute node. Crucially, despite these necessary variations in execution logic, the standardized configuration ensures that the benchmark itself remains identical. This allows us to generate comparable results across diverse environments.

\begin{figure*}
\begin{center}
\includegraphics[width=\textwidth]{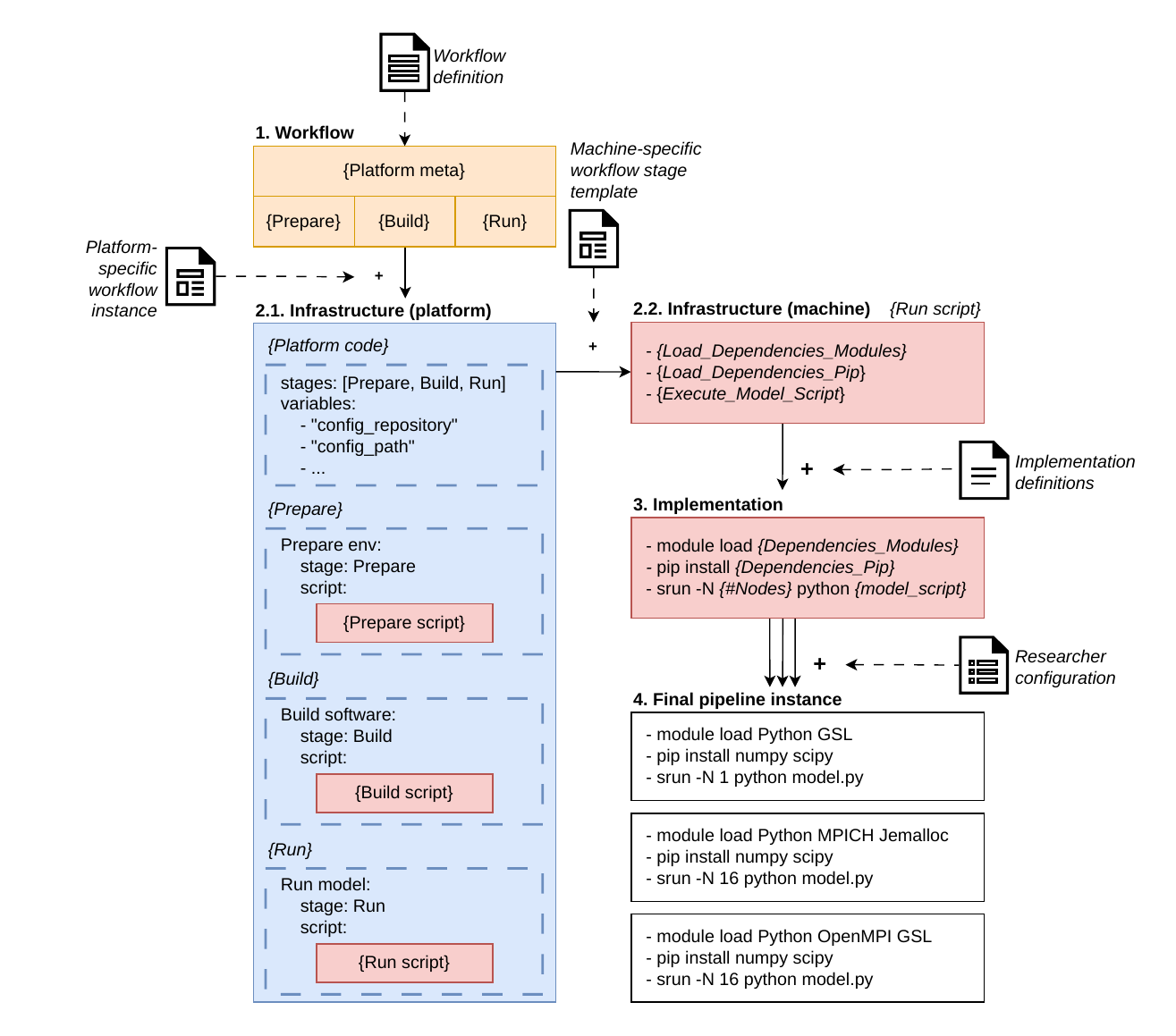}
\end{center}
 \caption{
   The template instantiation process starts with a workflow definition (1), specifying the individual stages of the resulting CI pipeline. The workflow template is specialized using platform-specific workflow stage instances tailored to the target platform using the corresponding syntax (2.1). These instances are machine-independent and are thus subsequently filled with machine-specific script template blocks for each workflow stage (2.2). These template blocks are next replaced by platform-independent command lists (3). Given the configuration provided by the researcher, a complete pipeline instance is created for each parameter combination in the provided parameter space by inserting the parameter values into the command templates (4).
 }
\label{fig:bench_concept_templates}
\end{figure*}

We propose a continuous benchmarking framework extending the concepts of \citet{Albers22_837549} to be implemented by research groups given their local requirements and conditions. We identify three layers of abstraction for each benchmarking pipeline. The first layer, the workflow layer, defines the overarching goal as a workflow, such as benchmarking, profiling, testing, or any other computational workload (for example large-scale simulation or data intensive applications). This layer remains research group- and platform-agnostic (making it independent not only of specific researchers but also of entire institutes). The second layer, called the architecture layer, details the steps required to execute each stage of the workflow on the designated platform and machine, accounting for research group specific expertise and standards. The layer can be logically split into two parts (platform and machine) for which a separate set of templates is created. The platform-specific templates serve as a skeleton of the CI pipeline using platform-specific formatting, like GitLab's YAML syntax. These templates specify the location where machine-specific code is subsequently inserted. Rather than defining explicit commands, the machine-specific templates function as high-level descriptions for groups of commands (such as ``load environment"). The architecture layer is solely concerned with workload execution and thus needs awareness of the targeted hardware, but remains unaware of specific benchmark details. Lastly, the implementation layer specifies the precise commands that must be executed within each step of the layer above. Its templates are both platform- and machine-agnostic. They are tightly coupled with the format of the benchmark configuration, which in a key-value-based format defines how to construct the benchmark run and provides the set of experiment parameters. Each layer defines a template which is successively specialized by each subsequent layer. The architecture and implementation layers additionally define a separate interface to the configuration. The result is a template for a single benchmark run which is subsequently fully specialized by a unique parameter combination when triggering the run (see \autoref{fig:bench_concept_templates}).

\begin{figure*}
\begin{center}
\includegraphics[width=\textwidth]{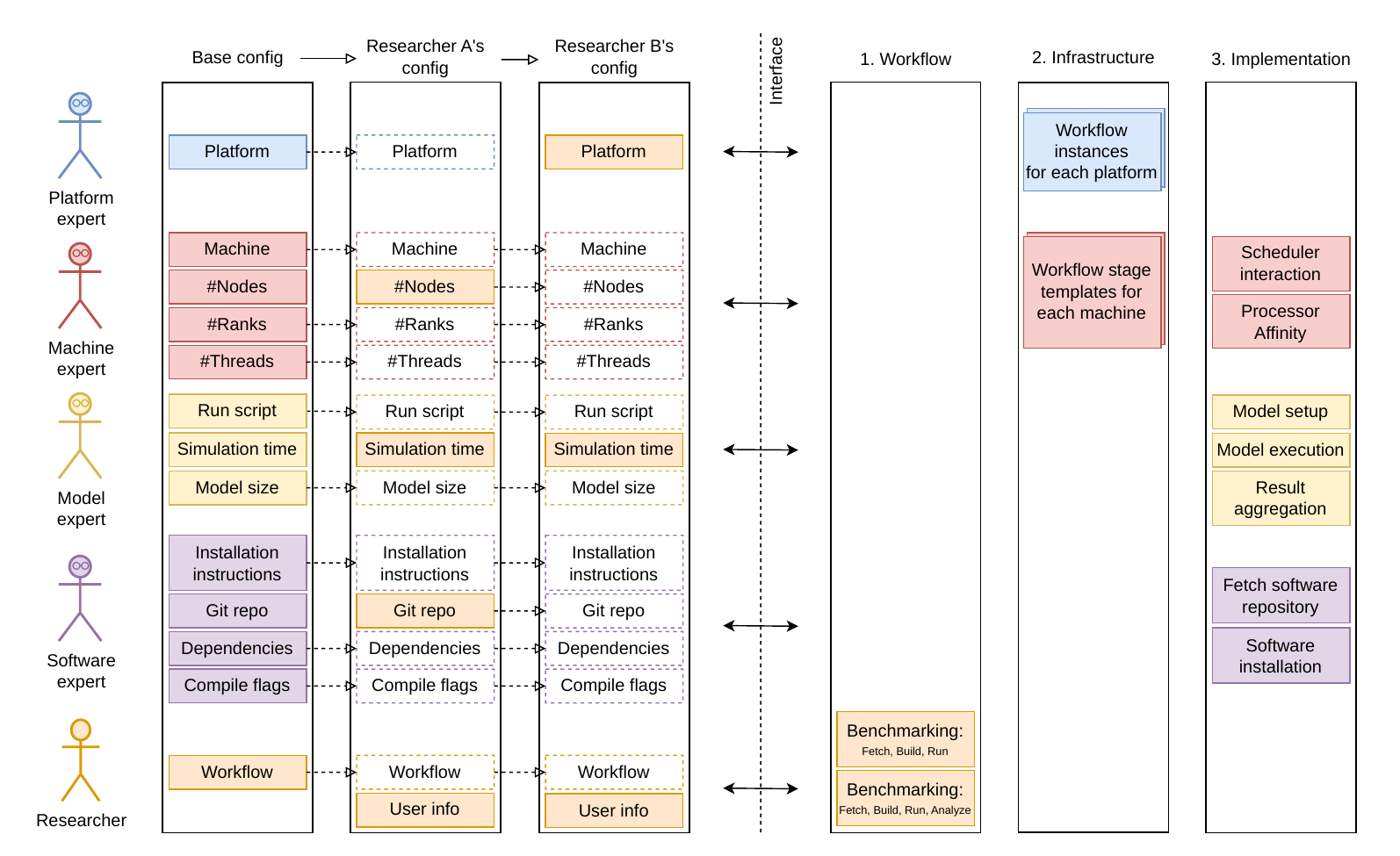}
\end{center}
 \caption{
    Division of responsibilities of configuration and template setup in a research group. Left: Each part of the base configuration for the group is authored by the corresponding expert. Individual researchers then create their own configurations by adapting parts of the configuration, either directly from the base configuration (Researcher A) or by building on configurations created by other researchers (Researcher B). A researcher's configuration extends either the base configuration or another researcher's configuration and only specifies the parameters to be overridden. The hierarchy of configurations always defines a fully specified configuration by inheriting values from parent configurations.
    Right: The configurable parameters are defined based on the corresponding templates created for each layer of the \texttt{CI-beNNch} framework (cf.~\autoref{fig:bench_concept_templates}).
    Color represents the expert or researcher authoring the configuration parameters or template implementations.
 }
\label{fig:bench_concept_roles}
\end{figure*}

By fully separating benchmark configuration and execution, the implementation details of the benchmarking process are adaptable and no longer up to the individual user: they can be addressed at a higher instance, e.g., by system administrators with expert knowledge about the HPC infrastructure. Providing a unified base configuration for the entire research group forms a framework which implicitly facilitates the division of responsibilities, where the different parameter sets can be added to the configuration by each corresponding expert, namely simulator setup, model configuration, and machine specifics. One can thus define five different roles: The \textit{machine expert} is responsible for the machine-specific code, such as the interaction with the HPC system's scheduler, as well as the interface to the machine configuration. The \textit{platform expert} integrates the machine-specific code into the general benchmarking framework and handles authentication of the CI runner to the machine. Both are infrastructure experts. The \textit{software expert} and \textit{model expert} define the software and model configuration, respectively. Lastly, the \textit{researcher} specifies the user configuration, which includes compute time accounting and project information, and adapts the base configuration to produce an experiment configuration (see \autoref{fig:bench_concept_roles}).


Each benchmark is thus fully specified by the hierarchical benchmark configurations together with the templates defined for each layer.

\paragraph{CI-based approach to benchmarking}
Based on the conceptual layers of abstraction, we define a controller which manages the execution of the benchmarking pipeline, by taking a configuration from the researcher at the unified entry point, constructing the benchmarking pipeline, and monitoring and adapting its execution (\autoref{fig:bench_concept}). Pipeline construction is divided into three parts: First, the controller generates an instance of a platform template by taking as input a workflow template, individual platform template pieces, and a platform configuration, producing a platform instance structured according to the targeted platform's format. Second, it produces a benchmark run script from this platform template and a given benchmark configuration. Certain commands are already explicitly defined, while others retain template variables that are subsequently filled with specific values provided by the benchmark configuration. This templated benchmark script can then be instantiated repeatedly, each time corresponding to a distinct combination from the parameter search space defined for the experiment in the benchmark configuration. Lastly, the controller is responsible for initiating benchmark runs using the generated run scripts.
The controller itself may be executed anywhere where it can be accessed by the group of researchers and is able to access the target machines of the benchmarks.

Consequently, a research group developing its own instance of the workflow is required to create the workflow, platform, and implementation templates given the choice of the CI platform and targeted benchmarking machines. Based on these templates, the configuration structure needs to be specified and a base configuration defined. Independently of the pipeline templates and configuration specification, the controller can be implemented on the system of choice, with the main requirement of being accessible to the whole research group.

\paragraph{Pipeline duplication}
In \autoref{fig:bench_concept_templates}, complete and independent pipelines are created for each parameter combination. In practice, this might lead to a lot of identical code being executed before the first parameter difference is encountered during pipeline execution. For example, if the \textit{build} stage is identical for all pipeline instances, it would not be necessary to perform this computationally expensive step several times.
Splitting up the pipeline into a sequential and parallel part can however often be non-trivial to implement, especially when trying to generically split the pipeline based on the input parameter space. However, experience has shown that in practice, it is typically sufficient to perform the split at the level of individual stages. Most of the common CI frameworks offer a simple and effective approach to parallelize individual stages of a pipeline and reuse the output of a sequentially executed stage via artifacts.

\section{\texttt{CI-beNNch} in action}
\label{sec:results}
Next, we explore specific, generally applicable implications of the design choices of the proposed continuous-benchmarking approach. We highlight advantages for commonly encountered situations in HPC benchmarking.

\subsection{Running a benchmark for the first time}
As noted above, the adoption of \texttt{beNNch} was primarily hindered by requiring an individual setup and configuration. Setting up and configuring \texttt{beNNch} is a laborious, error-prone, manual process and presents a high barrier of entry. In our experience, a successful use of \texttt{beNNch} usually requires the help of researchers already familiar with the tool as well as the peculiarities of the specific target machine.
The difficulty for a new researcher running an existing benchmark is now significantly reduced, assuming that other experts already set up a \texttt{CI-beNNch} instance including a base configuration. The CI-based approach introduces a higher degree of automation and eliminates the need for direct machine access altogether. Running a first benchmark with default parameter settings can now be reduced to two simple steps: First, the researcher executes a pipeline with default settings via the CI entry point. \texttt{CI-beNNch} then automatically builds the software environment, executes the benchmark jobs and gathers the results on the target machine. In the second step, once the benchmark jobs have been completed, the researcher is provided with the acquired data and visualizations of the benchmarking results. \autoref{tab:first_benchmark} provides a comparison of the setup steps required for \texttt{beNNch} versus \texttt{CI-beNNch}, highlighting the reduced complexity and improved accessibility for first-time users.

\begin{table*}
\caption{Initial execution of a benchmark using the \texttt{beNNch} framework versus a \texttt{CI-beNNch} pipeline. Red dots: manual user intervention. Blue dots: researcher must be logged onto the target machine. Black circles: steps automatically triggered with default settings but can be customized.}
    \begin{tabular}{m{2.5cm}|m{7cm}|m{7cm}}
       & \texttt{beNNch}  & \texttt{CI-beNNch}  \\ \hline \hline
    1. environment & \begin{itemize}
           \item[\textcolor{blue}{\textbullet}\textcolor{red}{\textbullet}] manually install software dependencies
         \end{itemize}
       & \\ \hline
    2. setup & \begin{itemize}
           \item[\textcolor{blue}{\textbullet}\textcolor{red}{\textbullet}] deploy \texttt{beNNch} (e.g. fetch, configure, and adapt machine-local instance, setup results location)
           \item[\textcolor{blue}{\textbullet}\textcolor{red}{\textbullet}] fetch models
         \end{itemize}
       & \begin{itemize}
           \item[\textcolor{red}{\textbullet}] get access to the CI entry point
         \end{itemize}
         \\ \hline
    3. configuration & \begin{itemize}
             \item[\textcolor{blue}{\textbullet}\textcolor{red}{\textbullet}] copy default benchmark configuration
         \end{itemize} &
         \begin{itemize}
             \item[\textcolor{black}{\textopenbullet}] research group provides base configuration which is used automatically
         \end{itemize} \\ \hline
    4. execution &
         \begin{itemize}
             \item[\textcolor{blue}{\textbullet}\textcolor{red}{\textbullet}] start benchmarking script
         \end{itemize} & 
         \begin{itemize}
             \item[\textcolor{red}{\textbullet}] trigger CI pipeline
         \end{itemize} \\ \hline
    5. analysis &
         \begin{itemize}
             \item[\textcolor{blue}{\textbullet}\textcolor{red}{\textbullet}] repeatedly check until all benchmarking jobs have finished
             \item[\textcolor{blue}{\textbullet}\textcolor{red}{\textbullet}] navigate to results folder on machine and execute analysis script
         \end{itemize} & 
         \begin{itemize}
             \item[\textcolor{black}{\textopenbullet}] analysis pipeline triggered automatically
         \end{itemize} \\ \hline
    \label{tab:first_benchmark}
    \end{tabular}
\end{table*}

\subsection{Reproducing a benchmark}
To overcome the limitations of \texttt{beNNch}, the \texttt{CI-beNNch} approach decouples benchmark configurations from individual researchers. All benchmark configurations are stored centrally and can be accessed by any researcher within the same research group. The automatic setup based on a centrally stored configuration ensures that the environment and procedures of reproduced benchmarks align as closely as possible with those of the original. The automatic generation of a standardized pipeline guarantees consistency across repeated benchmark executions, thereby facilitating repeatability and reproducibility and long-term reliability of research.

\subsection{Rerunning a benchmark with \dots}

\subsubsection*{\dots a different benchmarking model}
To comprehensively evaluate the performance of an HPC application, it must be tested across a range of different parameters. In the example of codes for the simulation of biological neural network, varying parameters may include biophysical mechanisms, connectivity patterns, activity dynamics, and network size. New benchmarking models are integrated into the benchmarking pool by updating a shared configuration, which specifies repository locations, model parameters, model version, and execution instructions. This ensures that once a model is added, it becomes immediately accessible to all researchers within the group.

\subsubsection*{\dots a different machine}
Comparative benchmarking across different machines allows researchers to identify whether performance is bound by the software itself or the underlying hardware. In the absence of an automated setup, following centrally defined machine-specific instructions, each researcher needs to individually set up the benchmarking environment on the target machine. As noted above,  creating these setups is time intensive, error-prone, and easily introduces differences hindering reproducibility. By contrast, in the proposed solution, the integration of a new target machine is done centrally for all users by adding additional machine-specific platform templates. After this initial setup, all researchers with access rights to the new machine can run benchmarks on it without further configuration.

\subsubsection*{\dots a different HPC application}
Adapting to a new HPC application involves specifying its deployment steps, required software dependencies, run instructions, as well as which performance metrics can be measured. Once this information is provided in the central configuration of the \texttt{CI-beNNch} instance, all researchers of the group are able to utilize the new simulation tool without any further adaptations.
In \texttt{beNNch} build environments and personalized modifications of \texttt{beNNch} are maintained individually by each user, often undocumented and inaccessible to others. This limits the reliability of shared build instructions and hinders reproducibility. In contrast, CI pipelines implicitly enable user-agnostic deployment and instructions which for the chosen simulation tool need to be verified only once.

\subsection{\texttt{CI-beNNch} in practice}
We developed a \texttt{CI-beNNch} instance based on GitLab CI/CD pipelines and Jacamar~\citep{jacamar}, an HPC focused CI/CD driver for GitLab which enables submitting jobs on HPC clusters. \texttt{CI-beNNch} has been implemented for the HPC systems of the Jülich Supercomputing Centre (JSC), namely JUSUF~\citep{jusuf2021}, JURECA~\citep{Thrnig21_1}, JUWELS~\citep{juwels2019}, and JUPITER~\citep{Jupiter} as well as a local compute cluster.

By defining a centralized entry point according to the principles outlined above, the entire complex process of correctly executing a benchmark is hidden from the researcher. We chose GitLab as a web-based entry point as it provides a straightforward user interface to trigger new benchmarking pipelines. The entry point pipeline is executed on a cloud runner, which downloads the configuration file provided by the researcher and constructs the combined benchmark specification by recursively loading and overwriting the hierarchy of configuration files up to the base level. Based on the final specification, a sub-pipeline is constructed for each target machine by instantiating and combining a set of pipeline templates.

For the task of benchmarking neural simulation codes, we define seven workflow stages which run as separate jobs in the sub-pipeline without any user intervention: Preparation, Build, Execution, Transfer, Annotation, Analyze, Plot. First, the environment is prepared by installing dependencies and fetching model and simulator code. This step requires internet access, restricting the choice of nodes of the system on which the job can be run (for example a login node or a developer partition). Second, the simulator is compiled on a compute node. This ensures an identical architecture for the subsequent benchmark runs, and the inclusion of micro-architecture specific optimizations without requiring a cross-compilation environment. The benchmark execution job itself runs on a login node, and submits compute jobs, one per parameter combination. Additionally, the raw data output from each compute job is enriched by a set of hardware and environment metadata from the compute node \citep{Villamar25_942}. The benchmark execution job waits for completion of all parallel compute jobs and aggregates the results of all benchmark runs. In the next stage, both the raw data and metadata are transferred to the final storage location. The collected data is subsequently analyzed (possible on any machine with access to the final storage location) by executing a default analysis routine which annotates the benchmarking data based on the recorded metadata. In this way, a standard for search and select mechanisms is provided. The final step produces a standardized graphical representation of the performance data and uploads it as an artifact to the pipeline. The user can view the generated plot through the same web interface as the entry point.

Beyond the use-case of CB, \texttt{CI-beNNch} is also suited for aiding feature development for HPC applications. There, similar problems arise when performance critical parts of the simulation code are touched: benchmarks should be performed across different scales on different machines. To highlight this use, we focus on the spiking neural network simulation engine \texttt{NEST} as an example \citep{Gewaltig_07_11204}.

\subsubsection{Barrier-free spike delivery}
In \texttt{NEST} version 2.16 the \textit{5g} simulation kernel has been introduced~\citep{Jordan18_2} which significantly improved the performance for large-scale simulations. Only a couple of years later, systematic benchmarking across scales revealed a drop in performance for specific cases, mainly due to newly introduced OpenMP barriers, thread-synchronization locations in the algorithm inevitably causing idle times of some threads. In \texttt{NEST} version 3.6 the \textit{5g} simulation kernel changes were combined with an optimized communication concept, where all spikes are communicated in a single buffer again to eliminate extensive thread synchronization (see \autoref{subsec:barrier-free}). The impact on performance is assessed across scales, with both weak and strong scaling, using the HPC-Benchmark model \citep{Kunkel14_78} (a scalable balanced random network), a microcircuit model \citep{Potjans14_785} (representing all layers of cortex with asynchronous irregular activity), and a multi-area model \citep{Schmidt18_e1006359} (a large-scale model of the macaque vision-related cortex) on the supercomputer JURECA-DC by comparing the last \texttt{NEST} version using the original \textit{5g} simulation kernel---\texttt{NEST} 3.5 \citep{nest35}---and the \texttt{NEST} release 3.8 \citep{nest38}.

\begin{figure*}[h]
\begin{center}
\includegraphics[width=0.75\textwidth]{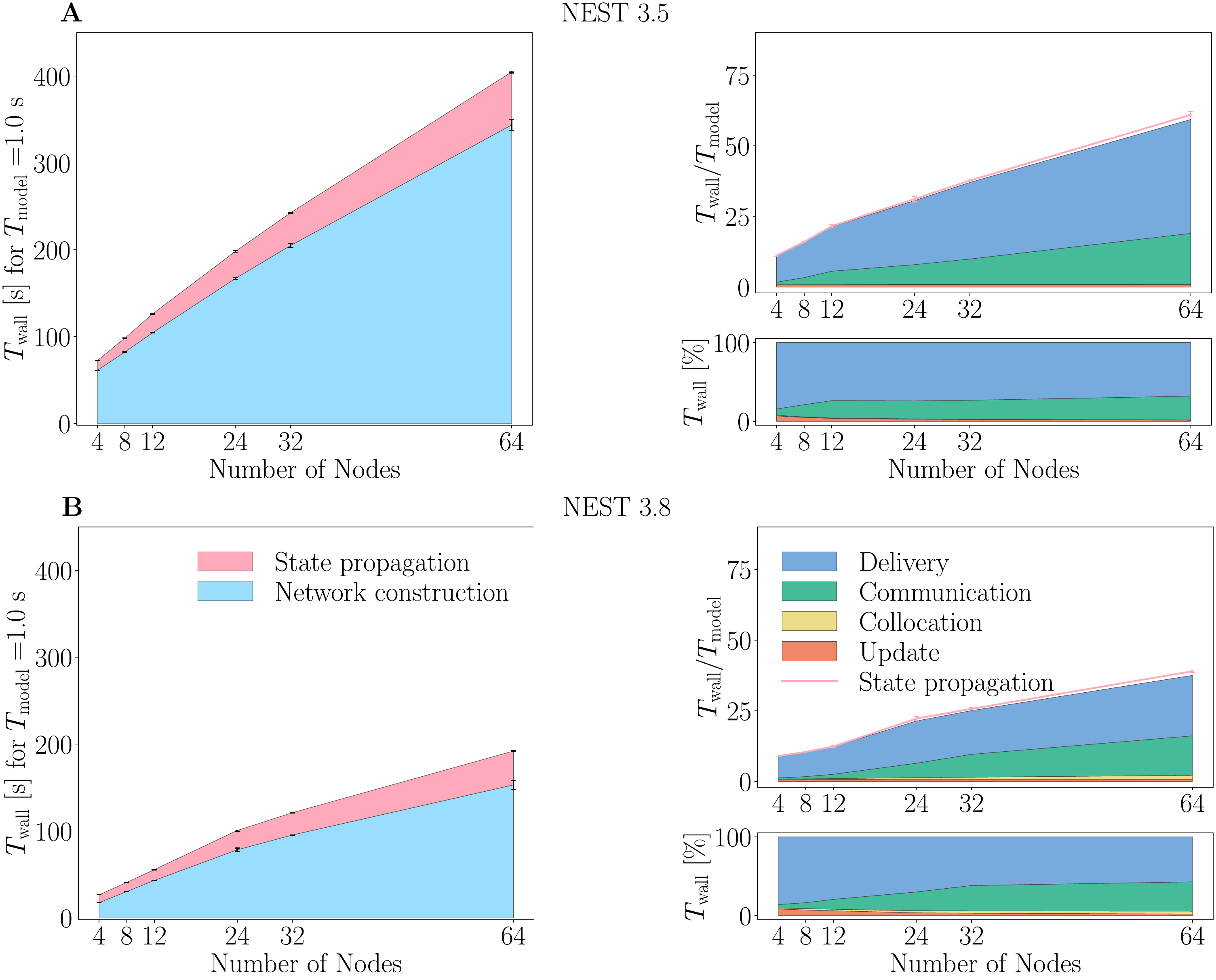}
\end{center}
 \caption{
 Weak-scaling performance of the HPC-Benchmark model on JURECA-DC using 2 MPI processes per node and 64 threads per process. (A) \texttt{NEST}~3.5. Left: absolute wall-clock time $T_\mathrm{wall}$ measured for both network construction and state propagation; simulating $90,000$ neurons and $10^9$ synapses per node for model time $T_\mathrm{model} = \SI{1}{\second}$. Error bars show standard error across three random seeds for each phase separately. Right, upper panel: real-time factor defined as wall-clock time divided by model time for each phase of state propagation. Error bars show the variability of total state propagation times across three random seeds. Right, lower panel: relative contribution of phases to the state-propagation time. See the legend of panel B for an explanation of the colors used in the figure. (B) \texttt{NEST}~3.8. Same display as (A).
 }
\label{fig:bench_HPC}
\end{figure*}

\begin{figure*}[h]
\begin{center}
 \includegraphics[width=0.75\textwidth]{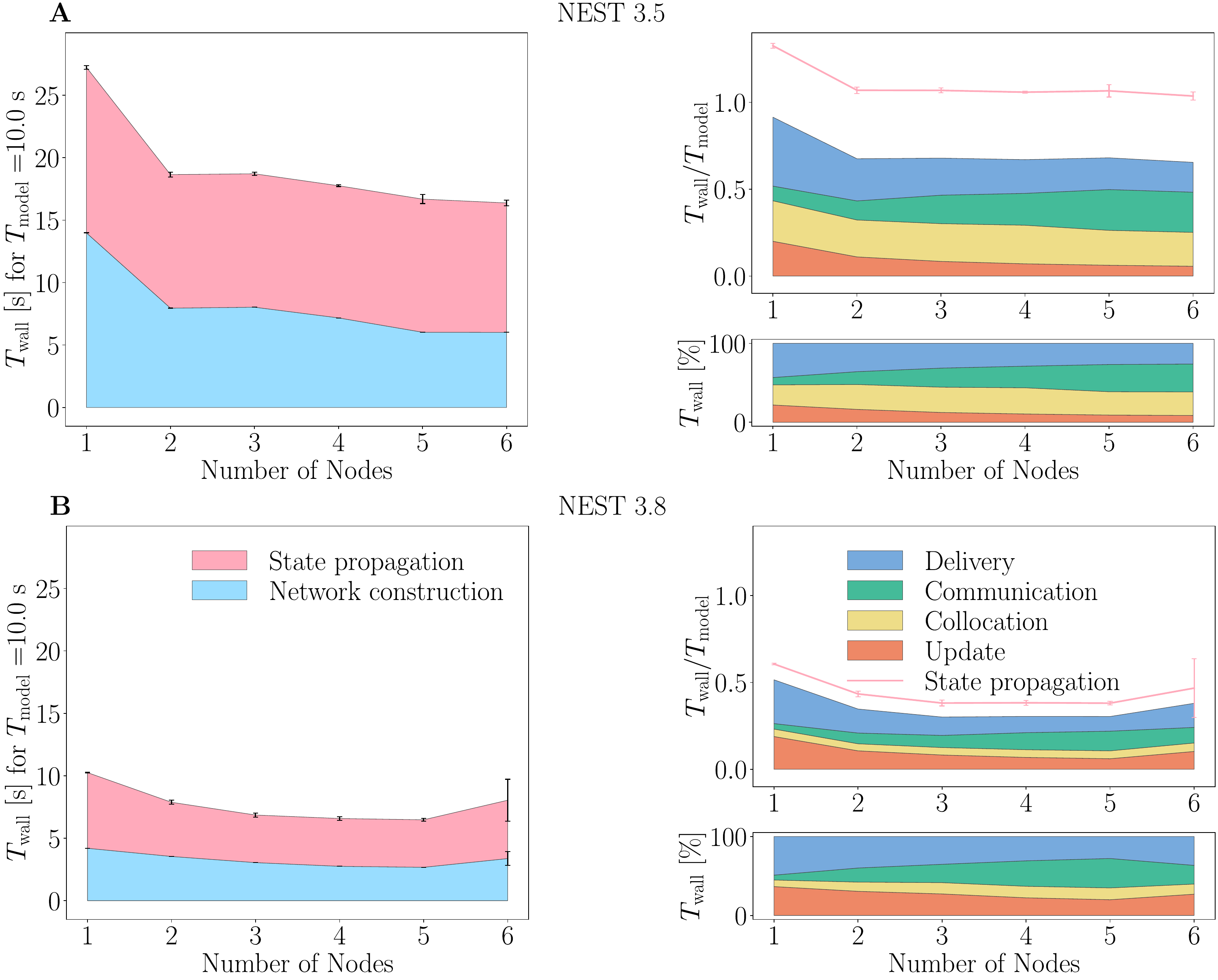}
\end{center}
 \caption{
 Strong-scaling performance of the microcircuit model on JURECA-DC using the same setup and display as in \autoref{fig:bench_HPC}. Model time $T_\mathrm{model} = \SI{10}{\second}$. (A) \texttt{NEST}~3.5. (B) \texttt{NEST}~3.8.
 }
\label{fig:bench_MC}
\end{figure*}

\begin{figure*}[h]
\begin{center}
 \includegraphics[width=0.75\textwidth]{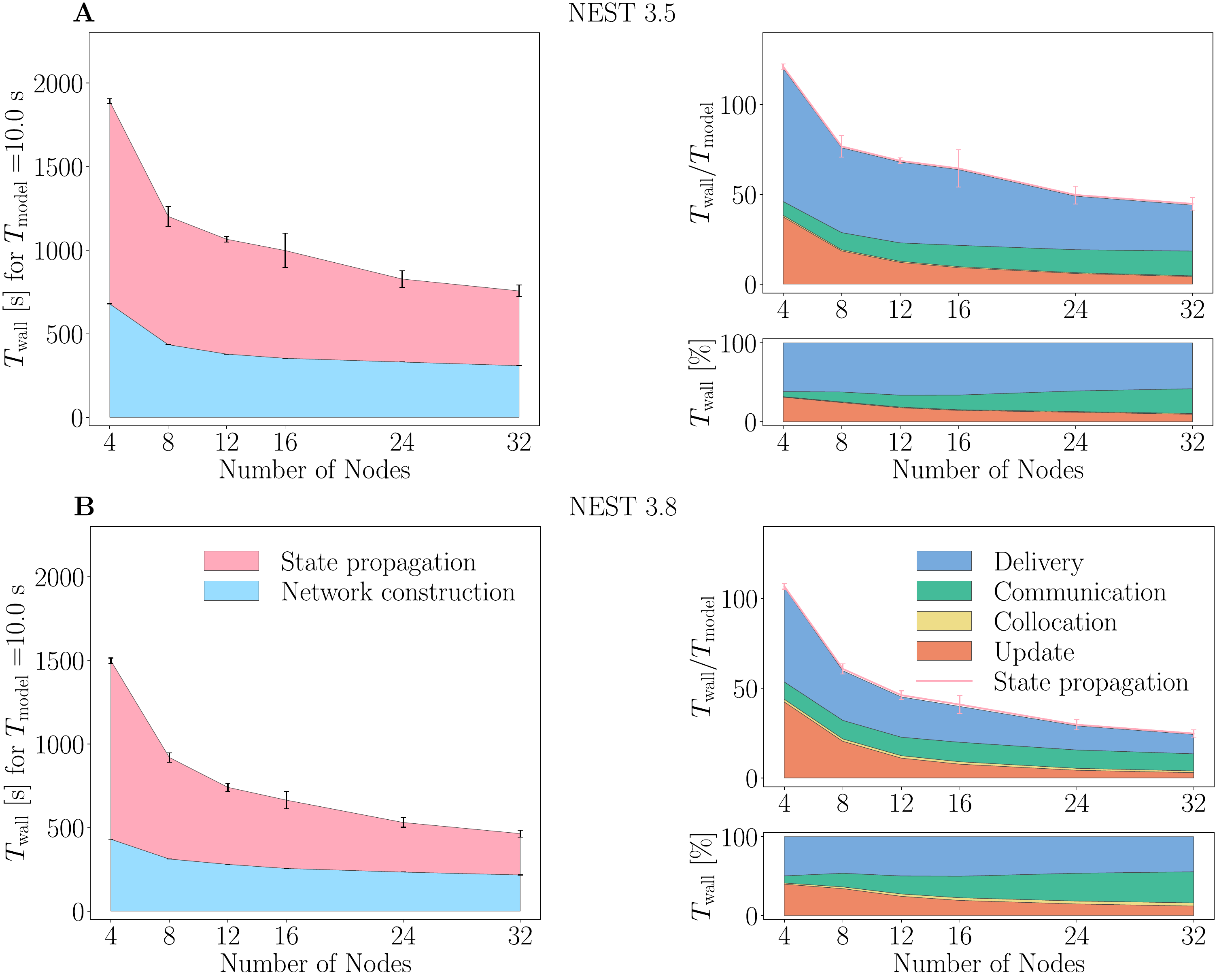}
\end{center}
 \caption{
 Strong-scaling performance of the multi-area model on JURECA-DC using the same setup and display as in \autoref{fig:bench_HPC}. The multi-area model is simulated in its meta-stable state for $T_\mathrm{model} = \SI{10}{\second}$. (A) \texttt{NEST}~3.5. (B) \texttt{NEST}~3.8.
 }
\label{fig:bench_MA}
\end{figure*}~

We distinguish four phases of state propagation: \emph{update}, \emph{collocation}, \emph{communication}, and \emph{delivery}. These concepts have proven useful in a sequence of earlier studies (for example \citep{vanAlbada21_47, pronold2022, pronold2022routing, Kurth_2022, Albers22_837549}): \emph{Update} measures the time required for updating the neuron dynamics. \emph{Collocation} means the time to collocate MPI send buffers from the spike register. \emph{Communication} refers to the time required to communicate spikes between computed nodes. And \emph{delivery} is the time needed to deliver spikes from the MPI receive buffer to the node local synaptic targets.

When comparing \texttt{NEST}~3.5 and 3.8, the benchmarks expose a reduction in spike delivery time across all models, leading to a consistent reduction of total simulation times across all scales. For the HPC-Benchmark, delivery times decrease by 45\% with a simulation time reduction of 26\% (\autoref{fig:bench_HPC}). Note that in comparison with \citep{Albers22_837549}, Fig. 4, the network construction time is increased due to altering the definition for the HPC-Benchmark model, making it consistent with the other benchmark models. The microcircuit model further benefits from simplified spike collocation (see \autoref{subsec:barrier-free}) reducing spike collocation by 76\% and spike delivery by 45\% for a total reduction of 60\% (\autoref{fig:bench_MC}). We observe substantial improvements for simulations of the large-scale multi-area model which complete 37\% faster (\autoref{fig:bench_MA}). While in \texttt{NEST}~3.5 simulations at larger scale are predominantly slowed down by spike delivery, with the changes introduced between \texttt{NEST}~3.5 and \texttt{NEST}~3.8, spike delivery becomes less of a bottleneck compared to spike communication when scaling up the number of compute nodes.

Network construction time is also markedly lower in \texttt{NEST}~3.8 than in \texttt{NEST}~3.5. The difference is particularly prominent for the weak scaling experiments with the HPC-Benchmark model (\autoref{fig:bench_HPC}), showing that network construction time has been reduced by an average of 60\%. This was achieved by a new approach to constructing the pre-synaptic connection infrastructure as detailed in \autoref{subsec:barrier-free}.

Changes to the simulation code were made by several researchers. A subgroup of them also was involved in assessing the performance of the suggested changes. \texttt{CI-beNNch} facilitated consistent benchmarking results between the individual researchers and thus contributed to rapid iterations in the development process.

\subsubsection{Speeding up spike-timing dependent plasticity}
One goal of Computational Neuroscience is to understand learning in humans and other mammals. Classical theories \citep{Hebb49} that are experimentally supported (starting from \citep{Bi98}) assert that a prerequisite to changes in synaptic strength between neurons (a hallmark signature of learning) is the synchronous activity of neurons, leading to computational models of spike-timing dependent plasticity (STDP).

When calculating updates to weights using STDP, a significant fraction of the update step is spent evaluating the exponential $\mathrm{exp}(-|\Delta t|/\tau_\pm)$, where $\Delta t$ is the elapsed time between post- and pre-synaptic spikes (assessing the ``coincidence'' of neural activity) and $\tau_\pm$ is a time constant. Since \texttt{NEST} employs a hybrid approach for state propagation, where events lie on a time grid defined by a fixed time step \citep{Morrison05}, we here assess to what extent pre-calculating exponential values and storing them in a look-up table improves performance. See \autoref{methods:exp} for more detailed considerations.

\begin{figure*}[h]
\begin{center}
 \includegraphics[width=0.75\textwidth]{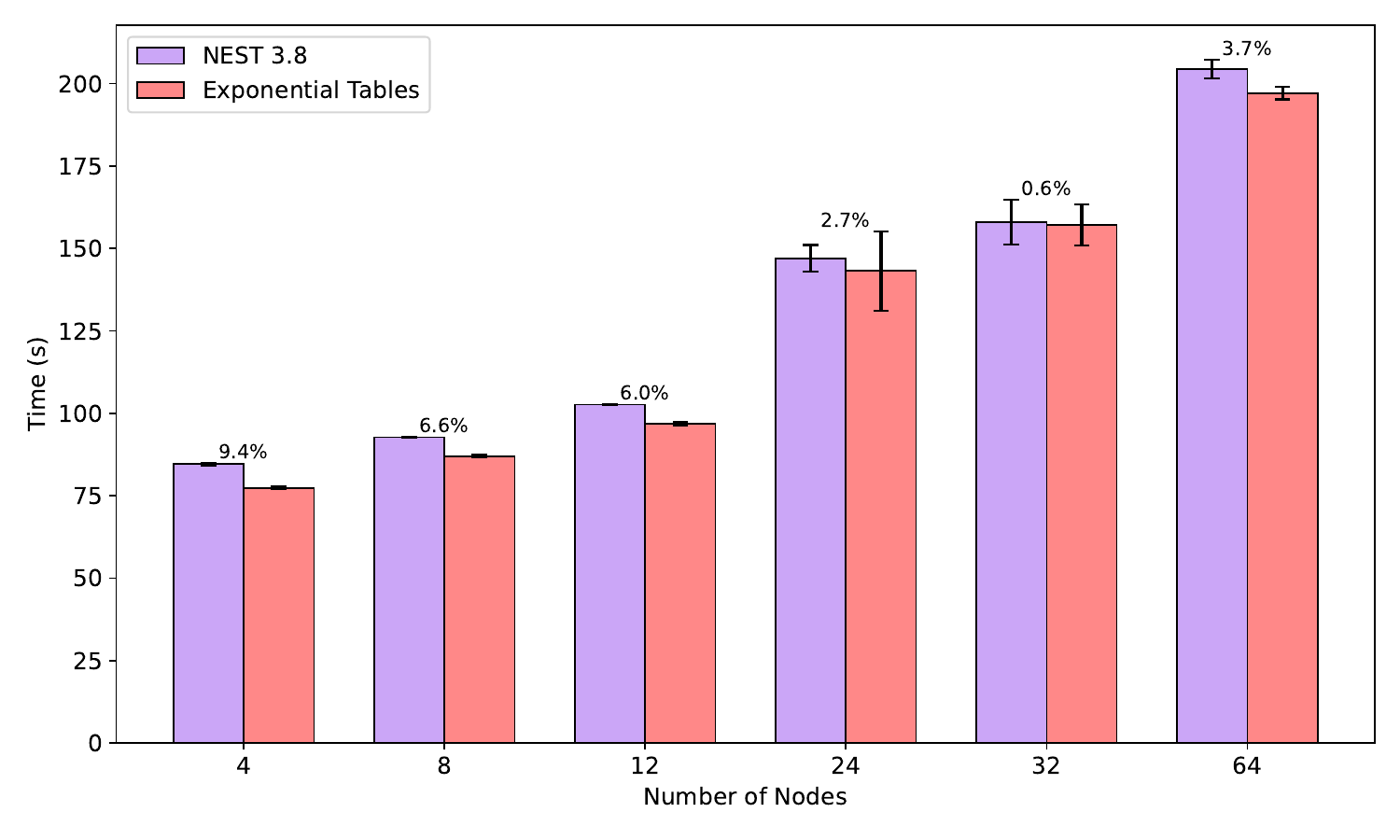}
\end{center}
 \caption{
 Differences in spike delivery time for a weak scaling of the HPC-Benchmark model on JURECA-DC using two processes per node and 64 threads per process between \texttt{NEST}~3.8 and the modified version using lookup tables for exponential operations. The model is run for $T_\mathrm{model} = \SI{10}{\second}$ with $90,000$ neurons and $10^9$ synapses simulated per node. Error bars show standard error across five random seeds, the percentage values represent the mean speedup of the exponential table implementation.
 }
\label{fig:exp}
\end{figure*}
We test this idea on the HPC-Benchmark against the \texttt{NEST}~3.8 release version, shown in \autoref{fig:exp}. We compare delivery time, since plasticity updates are performed during spike delivery in our reference code \citep{Morrison07_1437}. Lookup tables lead to an average reduction of about 5\% in delivery time across all scales. We observe an abrupt decrease of speedup at a high number of nodes ($64$). This is potentially due to increased computational background noise on the JURECA-DC system when utilizing a high number of nodes, including monitoring tools, or clogging the communication links.

Using \texttt{CI-beNNch}, it was possible to efficiently and swiftly test various intermediate implementations of the lookup tables by reducing overhead in the benchmarking process, allowing for greater focus on conceptual advancements and implementation details, as well as a division of labor between more and less experienced researchers.

\section{Discussion}
\label{sec:discussion}
In this work, we propose an extension of the benchmarking concepts developed in \cite{Albers22_837549} to unify and systematize the performance assessment of HPC applications. \citet{Albers22_837549} identified challenges regarding the reproducibility of performance assessments on a conceptual level. However, as experience showed, their approach for the practical execution of performance benchmarks---\texttt{beNNch} ---severely underestimates the remaining complexities in configuration, deployment, and execution of the HPC application and model code. Reproducibility of performance measurements, even for researchers working in the same laboratory, cannot in practice be ensured this way. We overcome these limitations by proposing a solution that abstracts configuration and deployment details. Our solution especially addresses issues regarding the transferability of concrete software setups and consequently improves reproducibility of results via an executor-agnostic approach. The advantages are threefold: the approach guarantees repeatability by avoiding laborious and error-prone individual set-ups; ensures reproducibility by providing global and inter-individual configurations specifying the environments, software deployment, and execution details; and enables replicability by allowing an easy transfer of details of the workflow between systems.

\subsection{Unified benchmarking executions}
Fundamental to the solution presented here is a common, machine-agnostic entry point from which all researchers of a research group can trigger benchmarks. This enforces sharing of configurations and enables identical executions. The steps required for benchmarking are carried out by a global controller, and thus do not depend on the individual researcher and their domain experience (see \citet{Pearce23} for a similar approach). While the approach of \citet{Anzt19_1} follows closely related principles, they do not highlight the possibility of achieving independence from target machines by a unified entry point.

\subsection{Hierarchical benchmark configurations}
To increase transferability and facilitate an executor-agnostic approach, we suggested the use of hierarchical benchmarking configurations. In this way, individual modifications of configurations required by researchers within the same group or from another research group can be addressed by adapting only certain parts of the benchmarking specification, while other parts remain identical. Hierarchical configuration facilitates division of labor between various experts who are responsible for the infrastructure, the HPC application, and the concrete benchmarks. In this way, the use is simplified especially for less experienced researchers, and thus efficiency and productivity of the benchmarking process can be increased. Defining clear responsibilities also increases transparency and thus aids the dissemination of knowledge. Our approach goes beyond \citet{Anzt19_1, Pearce23} who employ static configurations for the executed benchmarks. The flexibility stemming from the hierarchical configurations suggested here is particularly suited for the collaborative development of HPC applications, the performance of which is to be assessed across various systems.

\subsection{Centralized storage}
Comparing benchmarking results with data from the past is essential to continuously improve performance of HPC application. However, researchers often only store their results locally instead of centrally and data tends to get lost. Additionally, to repeat or reproduce benchmarks, the benchmarking configurations need to be saved so that they are accessible by other researchers within the same group and can potentially be shared with researchers from other groups. This is ensured by storing all configurations and any benchmarking data generated using these configurations in a centralized location. Enriching the data with metadata enables filtering for previously generated results and thus aids interpretation of new data \citep{Villamar25_942, Leipzig2021}. Additionally, the metadata is required for interpreting the previous data, as any changes in the experimental setup that might have an impact on the performance are recorded. Both aspects combined are required for continuous benchmarking.

\subsection{Limits of an executor-agnostic approach}
Due to security and accounting reasons, most HPC systems do not permit fully user-independent access to the system. A fully executor-agnostic approach to a benchmarking workflow is therefore usually not possible. In practice, this implies that user environments of individual researchers are automatically loaded when accessing the system. The issue of personal settings is in this sense entangled with accountability of the researcher. Some systems allow for the creation of additional user accounts that are not tied to a specific researcher, but the handling of corresponding compute budgets and authorization of access to resources becomes the responsibility of the account-providing researcher, making it impractical in many cases. This is, however, one solution to prevent leakage of user-dependent environment details, since in this case the loaded user environment is by definition consistent between individual researchers. Alternatively, the environment of the runner can be initialized empty via corresponding shell instructions. Similarly, containerization could be used to enable identical environments on any HPC system. However, for performance-critical applications, it is essential to utilize machine-specific modules to fulfill the software's dependencies, which can be difficult to achieve with general-purpose compiled code in containers. Instead, containers have to be carefully fine-tuned for each machine by system administrators, as building such high-performance containers is a highly non-trivial endeavor. However, this would come at the cost of flexibility of software dependency specification, as a specialized container would be required for each unique combination of dependencies, and system administrators could only offer a limited set of such combinations. Additionally, even for users with expert HPC knowledge, building such containers is oftentimes not possible. Due to proprietary soft- and hardware being employed on most HPC systems, building such containers is only possible on the HPC system itself, which makes it practically difficult or even impossible for non-priviliged users to trigger the build process or to access all required system and module settings, required for optimal performance. A weaker form of user independence is guaranteed in our approach by removing any researcher-dependence from benchmark configurations, while keeping accountability. This implies the researcher's responsibility to be aware of differences in the default environment of user accounts.

\subsection{Using \texttt{CI-beNNch}: A tale of Alice and Bob}
During performance benchmarks of a new feature of the \texttt{NEST} simulator, we observed unexpected behavior on one of the used systems: at the start of each simulation, there was an extended warm-up phase during which the performance was significantly reduced. Surprisingly, this warm-up phase was not apparent on another system with nearly identical hardware. Identifying the issue was a combined effort of multiple researchers spanning several months. Eventually, the cause could be identified as different \textit{NUMA-balancing} settings in the Linux kernel setting on the two systems.
To better understand this issue, multiple researchers executed benchmarks of the same model on two machines at different points in time.
\texttt{CI-beNNch} was used to great effect, guaranteeing the exact same benchmarking setup on both HPC systems. The issue could be investigated on the second machine immediately without requiring any additional setup, as the machine had been used for benchmarking already. The configuration for one machine could be reused on the second machine with only minor adaptations. The configurations for both machines were sharing the same parent configuration, which enabled us to centrally modify our experiments without any duplication of code or settings.
The time each involved researcher could allocate to this problem was frequently limited. Nonetheless, relevant experiments were performed by various subgroups without any overhead, as the configuration was centrally stored and accessible to all participating researchers. Specialized analysis of the benchmark results was necessary, requiring complex instrumentation of the simulation code. This led to a workflow of three experts working collaboratively on separate steps: simulator instrumentation, benchmark execution, and data analysis. Only minimal synchronous communication was required between the experts as these steps could be decoupled almost entirely by using \texttt{CI-beNNch}. After discovering the cause of the problem, a final report to the system administrators had to be created. This required gathering the code of all executed pipelines, the resulting data of each run, and the corresponding analysis figures. During this process, we noticed some shortcomings of our current \texttt{CI-beNNch} setup. The data for each benchmark, including both the benchmarking input and output, is centrally stored in an archive with a unique, auto-generated identifier. This avoids name collisions between archives, but makes it difficult to identify a benchmark executed a while back. For recent benchmarks, the timestamp or CI pipeline logs can be used to quickly find the data corresponding to the pipeline. These findings underscore the critical importance of robust data and metadata handling~\cite{Villamar25_942}.

\subsection{Conclusion}
Integrating successful optimizations, confirmed through systematic and automatized benchmarking, into production code requires substantial additional effort. This work should not be underestimated, as it falls outside the typical academic scope of a PhD project which usually focuses on developing and demonstrating new algorithms, and involves discussions within a larger group of experts. These challenges have only recently received widespread attention \citep{Hocquet24_1} and the field of Research Software Engineering (RSE) is forming to address them (see for example \cite{Speck24_8} or the BSSw initiative\footnote{Better Scientific Software Initiative: \url{https://bssw.io}}).

While the concepts developed in this study are targeted towards performance benchmarking, they also prove useful for application development. Mature scientific software projects often have multiple senior developers and frequently have contributions from less experienced researchers, may it be for educational or scientific purposes. \texttt{CI-beNNch} lowers barriers for rigorous testing and benchmark execution that would otherwise require time intensive onboarding of new contributors, and thus simplifies the development process as a whole. This additionally opens the possibility to use \texttt{CI-beNNch} in supporting researchers with limited HPC expert knowledge in the mathematical modeling of natural science applications, to transition models from local development environments to HPC systems \textemdash something that usually poses a significant barrier that can only be overcome with large time investments.

Taken together, by building on previous development \citep{Albers22_837549} and key insights from CI/CD, our solution conceptually advances the systematic as well as continuous benchmarking of HPC applications, and increases usability as well as reproducibility via unified entry points, hierarchical configurations, and depersonalization.

\section*{Conflict of Interest Statement}
The authors declare no competing interest.

\section*{Author Contributions}

Conceptualization: JVO, AK, DT, ML, CS, JVI
Methodology: JVO, DT, AK
Software: JVO, DT, ML
Validation: ML, JVO, CS, JVI
Formal analysis: ML, CS, JVO, JVI, HEP
Investigation: all authors
Writing - Original Draft: JVO, AK, ML, CS, JVI
Writing - Review \& Editing: all authors
Visualization: JVO, CS, ML, JVI
Supervision: AK, SK, MD, JS
Funding acquisition: MD, SK, HEP

\section*{Funding}
This project has received funding from
Volkswagen Foundation;
HiRSE\_PS, the Helmholtz Platform for Research Software Engineering - Preparatory Study, an innovation pool project of the Helmholtz Association;
the Joint Lab ``Supercomputing and Modeling for the Human Brain";
the European Union’s Horizon Europe Programme under the Specific Grant Agreement No. 101147319 (EBRAINS 2.0 Project);
Deutsche Forschungsgemeinschaft (DFG, German Research Foundation) – 545776403/FOR5880;
Juelich Research Centre intramural STEF fund for the update of instruments.
The Käte Hamburger Kolleg: Cultures of Research (co:/re) at RWTH Aachen supported Hans Ekkehard Plesser through a fellowship and provided financial, administrative and technical support for the workshop Building on Models using funding by the Federal Ministry of Education and Research (funding code 01UK2104).
Open access publication funded by the Deutsche Forschungsgemeinschaft (DFG, German Research Foundation) – 491111487.

\section*{Acknowledgments}
We thank the members of the NEST development community for their contributions to the concepts and implementation of the NEST simulator,
and our colleagues in the Simulation and Data Laboratory Neuroscience of the Jülich Supercomputing Centre for continuous collaboration.
We gratefully acknowledge the computing time granted by the JARA Vergabegremium and provided on the JARA Partition part of the supercomputer JURECA at Forschungszentrum Jülich (computation grant JINB33).
We acknowledge the use of Fenix Infrastructure resources, which are partially funded from the European Union's Horizon 2020 research and innovation programme through the ICEI project under the grant agreement No. 800858.
This project received access to the JEDI supercomputer, which is funded by the EuroHPC Joint Undertaking, the German Federal Ministry of Research, Technology and Space, and the Ministry of Culture and Science of the German state of North Rhine-Westphalia, through the JUPITER Research and Early Access Program (JUREAP).

\section*{Data Availability Statement}
The data and code are openly available on Zenodo\footnote{\href{https://doi.org/10.5281/zenodo.20133575}{https://doi.org/10.5281/zenodo.20133575}}.

\printbibliography

\clearpage
\onecolumn

\section{Supplementary Information}

\setcounter{section}{1}
\setcounter{figure}{0}

\renewcommand{\thesection}{SI \arabic{section}}
\renewcommand{\thefigure}{SI \arabic{figure}}

\renewcommand{\sectionautorefname}{SI}
\renewcommand{\figureautorefname}{SI Figure}

\label{sec:appendix}

\subsection{Use case: Barrier-free spike delivery}
\label{subsec:barrier-free}

Before the \textit{5g} simulation kernel, all spikes were always communicated in one go, independent of the number of spikes. This changed to multiple possible communication iterations with a fixed maximum communication buffer size to restrict memory consumption. After each communication round, all spikes are delivered to their respective local targets in parallel by all threads, after which a synchronization of all threads is required, before new spikes can be communicated again. In case of unequal delivery times, the required synchronization results in a significant amount of idle time. This synchronization was not needed before, when all spikes were communicated at once. The fixed buffer size has been particularly important for large systems with a high number of processes, as the communication buffer size increases linearly with the number of processes. At the time when the \textit{5g} simulation kernel was developed, the trend in supercomputer architectures showed an increasing number of compute nodes and therefore also an increasing number of processes (order of $100,000$) for large simulations. However, recent developments have shown an increasing popularity of fat compute nodes with many-core CPUs and an order of magnitude  lower numbers of processes (order of $10,000$) and high shared-memory parallelization instead.

As systems continuously evolve, so must the simulation code \cite{Aimone23_418}, which now no longer requires limiting the maximum communication buffer size anymore. Instead, an algorithm can now immediately communicate all spikes before starting delivery of spikes, thus eliminating the requirement for subsequent synchronization during spike delivery. In order to enable immediate communication of all spikes, it is necessary for all processes to know about the total number of spikes to be received. A straightforward approach would be to initially exchange the number of spikes each process intends to send globally, and then proceed with the second round of communication to transmit all spikes. However, empirical measurements demonstrate that on typical high-performance computing systems, network latency is the limiting factor rather than the available bandwidth. It is therefore advisable to avoid communicating twice, which effectively doubles the latency. Alternatively, based on the observation that the number of spikes communicated in different rounds is often similar and not changing rapidly, it may be assumed that the last used communication buffer size still suffices in the following communication period. The initial communication should be attempted, with the aim of sending all spikes, but if the first round shows that not all processes fit all their spikes in the communication buffer, the communication buffer is resized, and a second communication performed, which is then guaranteed to succeed. The present study implements and benchmarks this methodology. Additionally, the algorithm dynamically resizes communication buffers in accordance with current demand. If the communication buffer size needs to be increased, the buffer is expanded with a certain margin to account for a possibly further increasing number of spikes in the following rounds. To avoid using overly large buffers during periods of low network activity, communication buffers are shrunk when spike numbers fall below a given limit, again leaving a margin to allow for increased spike numbers. Shrinking does not require any additional MPI communication. Parameters controlling the growth and shrinking of buffers can be chosen by the user and statistics on resize operations are provided to the user to allow adjustment of the pertaining parameters to optimize performance.
Furthermore, parallel spike collocation is replaced by a simpler, sequential one, as it turns out to not be beneficial to parallelize this step even for highly parallel environments such as 64 compute nodes and 128 threads per node. For small models, sequential spike collocation has even shown to perform better compared to parallel collocation. In addition, the sequential code is easier to understand and thus to maintain.

Due to the complexity of the spike delivery process, profiling the code has shown that the compiler is not able to perform efficient loop unrolling when iterating over all spikes in the communication buffer. \cite{pronold_jari_2021_4564079} therefore introduce software loop unrolling by first extracting a fixed amount of spikes from the communication buffer, preparing all relevant data structures and only then performing batchwise processing of spikes by sending the spike to the corresponding synapse.

\begin{figure}[h]
\begin{center}
 \includegraphics[width=0.75\textwidth]{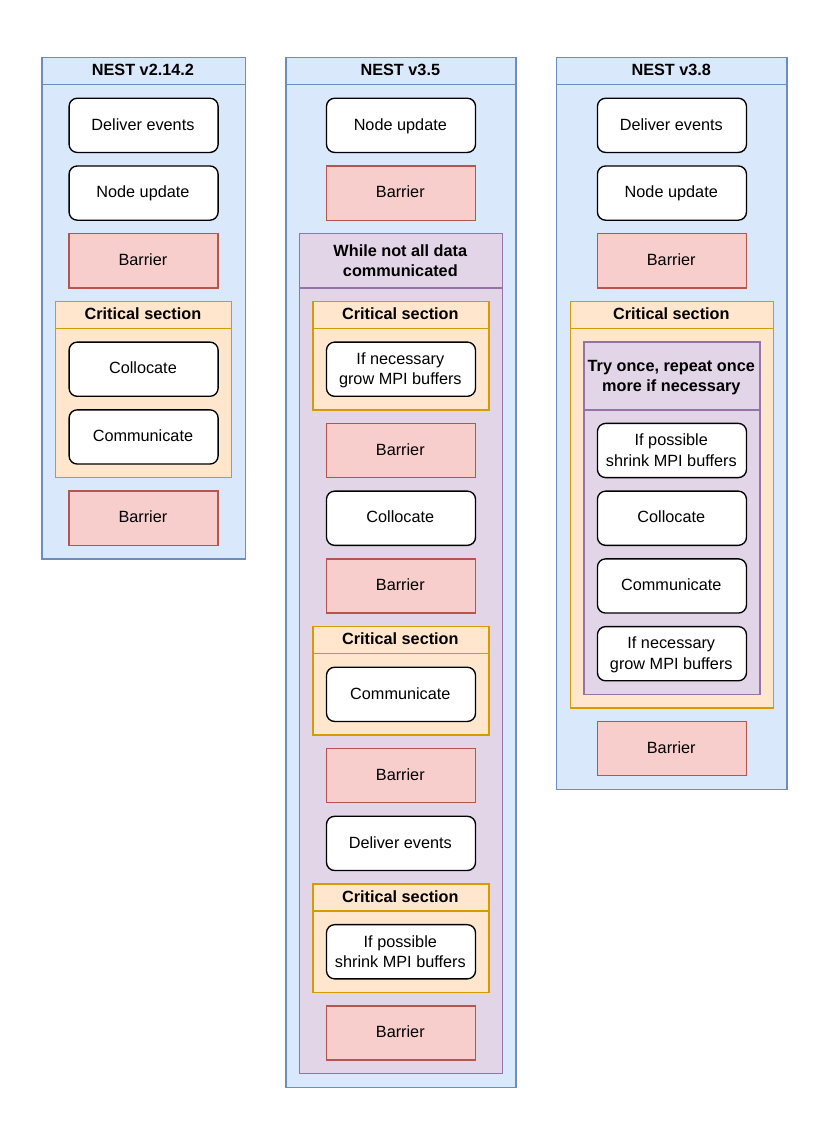}
\end{center}
 \caption{
 Simulation loop (blue) of \texttt{NEST} versions 2.14.2, 3.5, and 3.8. \emph{Deliver events} spans the logic to deliver spikes from the MPI receive buffer to the node local synaptic targets. During \emph{Node update} all neuron states are propagated to the next time step. In \emph{Collocate}, the spike register is collocated to the MPI send buffers. \emph{Communicate} refers to the process of exchanging spikes between computed nodes.
 Multi-threading barriers (red) indicate points of synchronization of all threads during parallel execution. The whole simulation loop runs in a thread-parallel mode, with the exception of critical sections, which are sequentially executed by the OpenMP master-thread only.
 }
\label{fig:sim_arch}
\end{figure}

Reference models are required to measure the impact of proposed improvements on simulation code such as a dynamic MPI buffer size. However, not all models are suitable to measure the performance of every simulation feature available.
This is particularly true for software aiming for scalability across multiple orders of magnitude regarding network sizes and computational demand, such as \texttt{NEST}. To properly measure the impact of MPI buffer size on simulation time, the instantiation of a large number of MPI processes with varying communication payloads is required. Previous studies~\citep{Jordan18_2,Albers22_837549} use the upscaled HPC-Benchmark and the macaque multi-area model~\citep{Schmidt18_e1006359} for this purpose and find an increase in simulation performance for large-scale models. Only after the publication of \citep{Jordan18_2} when the relevance of the code as a reference for neuromorphic hardware increased, researchers found a simultaneous degradation of performance for smaller models such as a cortical microcircuit~\citep{Potjans14_785}. In comparison to version 2.14, which demonstrates sub-realtime performance with a single MPI process~\citep{Kurth_2022}, more recent versions of \texttt{NEST} using the $5^{\textrm{th}}$ generation kernel do not match this single process performance, and require a decrease in the number of threads per process combined with a proportional increase in the total number of processes to reach optimal performance for the same computing resources. For quantitative analysis, we retrospectively add timers to version 2.14.2 including back-ported corrections as a reproducible reference. Along with further benchmarking and profiling analysis, it became clear that the communication loop required for dynamical MPI buffer sizes had a significant performance cost when using a large number of threads per process. To amend this, we move the spike delivery functions out of the communication loop, this brings the current architecture and performance closer to the previous version 2.14. \autoref{fig:sim_arch} illustrates the differences in the architecture of the main simulation loop for versions 2.14.2, 3.5, and 3.8 of the reference code. The figure exposes that version 3.5 requires more multi-threading barriers than 2.14.2. After refactoring, version 3.8 recovers the low number of synchronization points of 2.14.2 while still featuring a dynamic MPI buffer size.

At the outset of the investigation we struggled in identifying the critical differences between the different versions of the code. Due to the evolution of the code over a decade and the changing authors no diagram existed describing the different versions in the same terms. In addition, the ad hoc instrumentation of different versions by different researchers leaves doubts whether the reported run times for stages of the processing are comparable.
Based on earlier work \citep{Albers22_837549} the diagram (\autoref{fig:sim_arch}) now provides consistent terms across code versions for four main stages of the spike processing: deliver, update, collocate, and communicate. The detailed definitions are available in the \texttt{NEST} documentation\footnote{\url{https://nest-simulator.readthedocs.io/en/stable/nest_behavior/built-in_timers.html}} and are certainly subject to change as the code evolves. Furthermore, our work integrates the timers into the source code. Thus, the instrumentation is no longer ad hoc, but an integral part of the formal code review and any future release of the reference code. The need for such checkpoints accessible to every user and specified in the user-level documentation, is an aspect that distinguishes research software from an industrial application.

\subsection{Exponential lookup table for spike-timing dependent plasticity}
\label{methods:exp}

In spiking neural network simulations with spike-timing dependent plasticity, the pair-based update of synaptic weights $\Delta w^\pm\,=\,F_\pm(w)\cdot\mathrm{exp}(-|\Delta t|/\tau_\pm)$ with weight-dependency functions $F_\pm(w)$, temporal difference $\Delta t$ between post- and presynaptic spikes, and time constants $\tau_\pm$ consumes a significant amount of the run time in an otherwise latency-bound application (see \cite{Morrison08_459} for review). This is due to frequent computationally costly calculations of the exponential $\mathrm{exp}(-|\Delta t|/\tau_\pm)$. We therefore investigate whether the use of exponential lookup tables reduces simulation time even though the lookups potentially reduce cache efficiency.

As in traditional time-driven simulation codes for spiking networks, spike times are constrained to the time grid defined by the simulation resolution of typically $h\,=\,0.1\,\mathrm{ms}$, the temporal difference $\Delta t$ is also a multiple of $h$. This allows us to store precise, pre-calculated values $\mathrm{exp}(-|\Delta t|/\tau_\pm)$ in lookup tables for a range of temporal differences in steps of $h$, and only calculate the exponential during simulation if a requested specific temporal difference exceeds this range. Substituting or approximating expensive exponential calls is a well-researched practice \citep{schraudolph1999, Cawley00, malossi2015, DHaene-2009_1068}, leading modern neuromorphic chips to implement exponential and related functions directly in hardware \citep{partzsch2017}.

In order for exponential lookup tables to be applicable, potentiation $\tau_\mathrm{+}$ and depression constants $\tau_\mathrm{-}$ are required to be homogeneous across synapses, as table values represent the respective exponential of the scaled temporal difference. We initialize two tables, one for each time constant, with a size that represents a multiple of the resolution and is likely to cover the expected spike time differences, where generally a rather small range suffices. The exact size depends on the resolution as well as the dendritic delay in a given simulation. Should the former be set much lower than the standard \SI{0.1}{\milli \second} or in case of high extrema of the delay distribution, the table needs to cover more values. Running simulations with an increased resolution (i.e., smaller than \SI{0.1}{\milli \second}) is not a typical use case. Researchers use sample runs at higher resolution or with precise spike times~\cite{Morrison07_47, Hanuschkin10_113}, if available, to check the robustness of the simulation results.

The reference code represents spikes time as integer timestamps and floating point offsets, where the latter is only required in case of simulations with precise spike times. However, within the STDP class all time variables are floating point. This choice was made to support both grid-constrained and precise spike times. The lookup tables, however, require computations on discrete representations of time for efficient indexing. We therefore clone the original STDP code and modify the data types correspondingly. Furthermore, we also move $\tau_\mathrm{-}$ from the postsynaptic neuron to the shared STDP synapse properties for pre-computation.

During the process of integrating the prototype implementation into production code, our \texttt{CI-beNNch} setup enabled quick iteration cycles and ensured that no degradations of performance were introduced for applications not making use of the STDP-specific optimizations.

\end{document}